\documentclass[prd,preprintnumbers,twocolumn,amsmath,amssymb,nofootinbib,showpacs,secnumarabic]{revtex4}
\def\scn#1#2{\section{#1}\lb{#2}}
\def\sscn#1#2{{\it #1}.}

\usepackage{epsfig,amssymb,amsfonts,verbatim}
\def\bfl{\begin{flushleft}}
\def\efl{\end{flushleft}}
\def\bfr{\begin{flushright}}
\def\efr{\end{flushright}}
\def\bc{\begin{center}}
\def\ec{\end{center}}
\def\be{\begin{equation}}
\def\ee{\end{equation}}
\def\ba{\begin{eqnarray}}
\def\ea{\end{eqnarray}}
\def\baa#1{\begin{array}{#1}}
\def\eaa{\end{array}}
\def\bw{\begin{widetext}}
\def\ew{\end{widetext}}

\def\lb#1{\label{#1}}

\def\Sign#1{\, \text{sign}\left(#1\right) }

\newtheorem{Def}{Def.}[section]

\newtheorem{Prp}[Def]{Proposition}

\def\l{\lambda}

\def\cmo{\mu}
\def\l1{\eta}

\begin{document}

\preprint{Phys. Lett. A 375 (2011) 2305-2308  [arXiv:1003.0657]}

\title{
Vacuum Cherenkov effect 
%and other radiation phenomena\\
in logarithmic nonlinear quantum theory
}

\author{Konstantin G. Zloshchastiev}
%\email{Kostiantyn.Zloschastiev@wits.ac.za, bozons@gmail.com}
\affiliation{
%National Institute for Theoretical Physics (NITheP),
%Stellenbosch Institute for Advanced Study,
%Stellenbosch, South Africa}
%\affiliation{and Institute of Theoretical Physics, University of Stellenbosch, Stellenbosch 7600, South Africa}
%\affiliation{
Department of Physics and Center for Theoretical Physics,
University of the Witwatersrand,
Wits 2050, Johannesburg, South Africa
}

%\affiliation{Department of Physics, National University of Singapore,
%Singapore 117542 %, Republic of Singapore
%}

%\affiliation{Department of Theoretical Physics, Dnepropetrovsk National University, Dnepropetrovsk 49050, Ukraine}

%\date{~ ~~~~~~~~~~~~~~~~~~~~~~}
%\date{Received: 2 Mar 2010}
%\date{~Received: 26 May 2000 [PRL], 1 June 2000 [LANL] ~}
%\date{Received \today}

%\scriptsize%\footnotesize

\begin{abstract}
We describe  the radiation phenomena which
can take place in the physical vacuum 
such as
Cherenkov-type shock waves.
Their macroscopical characteristics - cone angle, flash duration,
radiation yield
and spectral distribution -
are computed.
It turns out that the radiation yield
is proportional to the square of the proper energy scale of the vacuum
which serves also as the vacuum instability
threshold and the natural ultraviolet cutoff.
While the analysis is mainly based on the theory engaging 
the logarithmic nonlinear quantum wave equation,
some of the obtained results 
must be valid
for any Lorentz-invariance-violating theory 
describing the 
vacuum by (effectively) continuous medium in the
long-wavelength approximation.

%~\\ \textit{Key words}: Quantum gravity - elementary particles - beyond relativity - gamma rays: bursts
\end{abstract}


%\keywords{Cherenkov radiation, non-trivial vacuum, logarithmic Schroedinger equation, Lorentz invariance violation, physical vacuum models}
\maketitle

%\narrowtext

%\normalsize
%\newpage

\scn{Introduction}{sec-i}

Current 
observational data in astrophysics seem to indicate the existence
of the deviations from the classical theory of relativity \cite{Kifune:1999ex,Protheroe:2000hp,AmelinoCamelia:2001qf,Stecker:2009hj,Thulasidas:2007hv,Chang:2008zzr}.
In absence of a fully satisfactory axiomatic theory explaining them, 
there appeared numerous non-axiomatic theories and 
effective approaches 
broadly referred as the effective quantum gravity theories.
One of the candidate theories has been proposed in \cite{Zloshchastiev:2009zw}
based on the 
nonlinear logarithmic quantum mechanics \cite{BialynickiBirula:1976zp}
which is still a  subject of intensive study nowadays, in particular,
in the connection with the quantum locality issues \cite{epr}.
%conjecture that the physical vacuum causes the universal deformation of quantum wave equations making them nonlinear, of the  type 
The idea was alternatively formulated on the field-theoretical language in the subsequent paper
\cite{Zloshchastiev:2009aw}.
There 
%it was shown that 
the 
necessity
of introducing the
universal nonlinearity in the quantum wave equation
%for the Universe's wave function 
was
explained by the arguments that
the physical vacuum is a kind of the non-removable background Bose liquid
or Bose-Einstein condensate (BEC)
located in a fictitious Euclidean space.
According to the superfluid vacuum approach,
Lorentz symmetry is an emergent low-energy phenomenon
(on a vacuum energy scale),
and
the Standard-Model particles and gravity can be treated as the small 
fluctuations of the non-relativistic background superfluid  \cite{Sinha:1976dw,Novello:2002qg}.

%\newpage

%\scn{``Luminal boom'' and shock waves in vacuum}{sec-exp}

As long as we take the point of view that
the physical vacuum can be described as the
nontrivial medium which is continuous in the long-wavelength
approximation,
one of the predictions of the phenomenological 
approach \cite{Zloshchastiev:2009zw}
comes by analogy with the (Vavilov-)Cherenkov effect\footnote{In  literature 
the distinct 
transliteration, \v Cerenkov, is  often used. It is actually
a misnomer because
checked chars exist neither in  Russian nor in Latin alphabets.}.
The latter
is known to happen in the conventional
materials 
%such as the water or ice 
because the phase velocity of light in those media is less than the fundamental velocity $c$, and 
%thus in such media high-energy 
particles of non-zero rest mass
can propagate faster than photons \cite{cher,Frank:1937fk}. 
Unlike other associated radiation phenomena, such as \textit{Bremsstrahlung},
Cherenkov radiation is the collective response of the whole medium which is
essentially universal (in particular, material-independent), polarized
and directed along the beam, also its spectrum is continuous  with maximum
of intensity shifted to the higher-frequency (``ultraviolet'') side.
%How about the possibility of the ``luminal boom'' and Cherenkov shocks in the physical vacuum alone?

\scn{Deformed dispersion relations}{sec-ddr}

The classical theory of relativity clearly forbids any superluminal phenomena \cite{Mignani:1973pq}: 
a particle with non-zero
rest mass can reach speed of light only at infinite energy
(besides, the nontrivial vacuum itself would create a preferred frame of
reference, in violation of one of the relativistic postulates).
On the contrary, in the Lorentz-invariance violating (LIV) theories
the dispersion relations
alter hence the vacuum Cherenkov radiation is  not excluded indeed \cite{Beall:1970rw,Bolotovskii:1972ve,Coleman:1997xq,jlm2003,Lehnert:2004hq,Castorina:2004hv,Kaufhold:2005vj,Altschul:2006zz}.
For instance, in the theory \cite{Zloshchastiev:2009zw}
the (velocity) dispersion relation
for the particles approaching the speed-of-light barrier
is derived as  
\be\lb{e-vnpt}
v/c
=
\left[
1+
\cmo
\left(
1-
\frac{E}{E_0}
\right)^2
\right]^{-1/2}
,
%\ v \sim c ,
\ee
where 
$\cmo  \ll 1$
is the emerging effective parameter
which
by construction does not depend on energy  of
a particle but may vary for different species of particles
%(it can be interpreted as a constant of refraction, see below),
$E$ is the energy of a particle,
and
$E_0$ is the proper energy of vacuum (we remind that it is not 
necessary positive).\footnote{We emphasize that this dispersion relation
is valid in the vicinity $v \approx c $ and for non-zero $\mu$ only.
In the classical limit the vacuum becomes trivial, $\mu$ must be set
to zero (since its origin is essentially quantum), and the relation turns into an identity.}
Thus, below we shall
describe a formal experiment where assume that physical
particles (including photons) obey dispersion relations of the type (\ref{e-vnpt}),
differing only by the values of the parameters $\mu$.

Before proceeding any further, we define what is meant by the speed of light in our case.
The old definition, the fundamental constant $c = 299792458$ m/s, becomes now a 
unit conversion number which
refers
to the (approximate) speed of the photon whose energy is very small
comparing to $|E_0|$ - because this is how this speed was measured in past.
The genuine physical velocity of photons $c_n$ is energy-dependent
and determined by Eq. (\ref{e-vnpt}),
\be\lb{e-vnpt-ph}
c_n/c
=
\left[
1+
\cmo_\gamma 
\left(
1-
\frac{E_\gamma }{E_0}
\right)^2
\right]^{-1/2}
,
%\ v \sim c ,
\ee
where $E_\gamma $ is the photon's energy,
and the
parameter $\cmo_\gamma $ must be positive for the 
restriction $c_n \leqslant c$ to hold.\footnote{However, in principle the BEC vacuum is a dynamical medium, therefore,
one can not exclude the possibility that  $\mu_\gamma$ can turn negative
in some physical
situations. 
Then $c_n > c$ and for an accelerating charged particle it is the
vacuum instability which would appear first,
the subsequent appearance of the sole electromagnetic
Cherenkov wave seems unlikely.}
%Later we  reformulate $c_n$ in terms of wave frequency and refractive index $n$.
From these dispersions one can immediately deduce that the  behavior of the velocity
of a particle in the vicinity of either of speed-of-light points
is regular.
This essentially means that the ``barriers''
corresponding to the physical speed of light $c_n = c/n$ and 
fundamental velocity $c$
are not infinite anymore
and the velocity of the particle of sufficiently high yet finite
energy can reach these
values exactly.
For the classical particle the required energy scale would be about $E_0$
which can take any value  up to the Planck scale.
However,
%as long as the speed barrier is finite 
quantum particles can penetrate these barriers at lower energies,
and
the probability of this grows along with the particles' energy and velocity.
According to the above-mentioned BEC interpretation,
$E_0$ is a  critical scale at which
the fundamental 
background becomes unstable against the phase transition where
the physical degrees of freedom alter.
In practice this instability should most probably cause the quantum many-body effects, 
such as
the vacuum polarization and  creation of pairs, which would drain energy away.
In that case
we do not have a
single-particle problem anymore, 
also we cannot  neglect the
particle's velocity change.

As long as the primary target of our paper is the Cherenkov
radiation,
in what follows by the speed-of-light barrier
we will understand the $c_n$ one, all particles are assumed
to be 
in the subluminal mode, $c_n \leqslant v \leqslant c$,
as defined in Ref. \cite{Zloshchastiev:2009zw}.
As for the superluminal particle which crosses the $c_n$-barrier while decelerating 
then the physical picture is not clear yet, as one should deal
with the problem in an essentially 
many-body way \cite{Unruh:1976db}.

To proceed with derivation of
the properties of the Cherenkov radiation 
we first recall
that among all  nonlinear extensions of quantum mechanics
it is
only
the logarithmic one which jointly conserves two physically very important
features of the conventional (linear) quantum mechanics.
The first property is the additivity of energy 
for uncorrelated systems,
and
the second one is that
the Planck relation,
$
E = \hbar \omega
,
$
holds for stationary states \cite{BialynickiBirula:1976zp}.
%These properties will greatly facilitate our life afterwards.
Using the latter and Eq. (\ref{e-vnpt-ph})
one can immediately write down an expression
for the effective refractive index 
in the Cauchy form
%for photons
\be\lb{e-refi}
n^2 = 1 + \cmo_\gamma 
\left[1+ {\cal M}(\omega) (\omega/2 \pi c)^2
\right]
,
\ee
where 
the parameter
$
\cmo_\gamma 
%=  \chi_\gamma^2 -1 \ll 1
$
can be thus  interpreted  as indeed the constant of
refraction of the physical vacuum,
$
%\be\lb{e-disp}
{\cal M} (\omega)
=
\frac{4 \pi^2 c^2}{\omega_0^2}
\left(
1
+
2 \xi
\frac{\omega_0}{\omega}
\right)
$
is the effective dispersion coefficient,
$\omega \equiv E_\gamma /\hbar$ is the angular frequency of the electromagnetic wave,
$\omega_0 = |E_0|/\hbar$ is the natural  frequency of the vacuum,
and $\xi = -\Sign{E_0}$.
All this means that both the elementary particles and
electromagnetic waves propagating
through the physical vacuum get affected by it,
and once again confirms that the physical vacuum is a medium with non-trivial
properties.

As long as the Cherenkov effect is an essentially macroscopic long-wavelength
phenomenon
(the particles' Compton wavelengths 
are much larger than the characteristic size $\ell_0 = h c/|E_0|$),
its main properties 
can be easily computed just using the dispersion relations above.
%from the logarithmic theory.

\scn{Cherenkov cone angle}{sec-can}

We consider the following physical setup:
a particle moving with speed $v$, $c_n \leqslant v \leqslant c$,
% (very close to speed of light), 
momentum $\textbf{p}$
and energy $E$ emits at some point the photon with energy $E_\gamma$,
momentum $\textbf{p}_\gamma$
%frequency $\omega_\gamma $
and velocity $c_n$.
After this event the particle acquires 
speed $v'$, momentum $\textbf{p}'$
and energy $E'$.
Before and after the moment of emission
the particles are assumed to be uncorrelated.
Then, due to the above-mentioned energy additivity property 
we can write
the energy conservation law in the standard form,
%\[ (E-E_0) = (E'-E_0) + (E_\gamma - E_0) + E_0 , \] which yields
$
E' = E - E_\gamma
,
$
where the photon energy can be expressed as
$
E_\gamma = h c_n/\lambda
$
with $\lambda$ being the photon wavelength.
%Apparently, we may not have the naive additivity property 
For momenta
we can write
the standard conservation law as well
if
we work in the reference frame
where the momentum of the background is set to zero,
and
obtain the standard expression
for the conical angle:
%\be\lb{e-cos}
$
\cos \theta
=
1 - 
\frac{{p'}^2 - 
{\left( p - p_\gamma \right) }^2}{2\,p\, p_\gamma}
$
provided 
$p < p' + p_\gamma$.

Further, for future it is convenient
to introduce the following
dimensionless quantities
$
M = v/c,\
M' = v'/c,\
M_\gamma = c_n/c = 1/n,
$ 
$
%\nn\\&&
\epsilon = E/E_0, \
\epsilon' = E'/E_0, \
\epsilon_c = ( h c/\lambda)/E_0, 
$
$
%\nn\\&&
\epsilon_\gamma = E_\gamma/E_0 = ( h c_n/\lambda)/E_0, \
\sigma = \Lambda / \lambda = (h/p)/ \lambda,
%\nn\ea
$
and their combinations 
such as the inverse Lorentz factors
$\Gamma = \sqrt{1 - M^{2}}/M$,
$\Gamma'$, $\Gamma_\gamma$, \textit{etc.}
%It should be remembered that the epsilon values are not necessary positive.
Then the velocity dispersions
for our setup can be written
as
$
M
=
1/\sqrt{
1+
\cmo
\left(
1-
\epsilon
\right)^2
}
,
$
$
M'
=
1/\sqrt{
1+
\cmo'
\left(
1-
\epsilon'
\right)^2
}
%\\ &&
$
and
$
n 
=
\sqrt{
1+
\cmo_\gamma
\left(
1-
\epsilon_\gamma
\right)^2
}
.
$
Further, as long as $M$ and $M'$ refer
to the same particle in similar physical condition
we must impose
$
\cmo' = \cmo,
$
from which we obtain 
the
velocity transformation formula
\be
M'
=
\left[
1+
\Gamma^2
\left(
\frac{1-\epsilon'}{1-\epsilon}
\right)^2
\right]^{-1/2}
,
\ee
which can be used to eliminate $M'$ where necessary.
Then,
using the momentum dispersion relations
one can write the cone angle  in the
form
\be
\cos \theta
=
1 - 
\frac{P^2(\epsilon - \epsilon_\gamma,
\Gamma
\frac{1-\epsilon + \epsilon_\gamma}{1-\epsilon})
- 
{\left( p - P(\epsilon_\gamma,\Gamma_\gamma) \right) }^2}{2 p P(\epsilon_\gamma,\Gamma_\gamma)}
,
\ee
where 
%by $P (\epsilon, \Gamma)$ 
we defined the function  
%\bw
$
P (x, y) 
\equiv
\frac{E_0 (1-x)}{2 c y}
%\left(1- \frac{E}{E_0}\right)
\biggl[
\Upsilon
\left(
\frac{y}{
1-x}
\right)
-
\Upsilon
\left(
y
\right)
\biggr]
$
assuming
$ \Upsilon (x) \equiv  x \sqrt{1+x^2} + \text{arcsinh}\, x$.

%\ew
Using the formulae written above,
it is straightforward to write down the 
exact expression for the cone angle
as a function of $M$, $n$, $\sigma$ and $\epsilon_c$.
In general, this expression is quite bulky (in particular,
it involves
the solving of the transcendental equation 
%to obtain the dependence 
for $\epsilon (p)$ if one wants to obtain an expression in terms
of de Broglie's wavelengths)
and thus it is suitable more
for a numerical analysis.
For analytical purposes,
one can write it in a perturbative form,
using the smallness of the parameters
$\l1 = 1 - M^2$ and
$\l1_\gamma = 1 - M_\gamma^2 = (n^2 - 1)/n^2$.
% $\epsilon_c$ and $\epsilon_c / \sigma$.
This approximation is valid
as long as all the velocities in the problem
are close to $c$.
%and the characteristic frequency of the emitted Cherenkov photon is small comparing to the natural frequency of the physical vacuum.
We obtain
\be
\cos \theta
=
\frac{1}{M n} 
+ 
\Theta_q
,
\ee
where $n$ is given 
by Eq. (\ref{e-refi}),
and
by $\Theta_q$ 
we denote the correction term
\bw
\be
%&&
\Theta_q
=
\frac{1}{2} \l1_\gamma
 \sigma
+
\frac{1}{3}
\l1_\gamma
 \frac{\epsilon_c
        ( \epsilon_c  - 3/2)(1-\sigma)   
       }{\left( 1 - {\epsilon_c } \right)^2}
 + 
 %\nn\\&&\qquad \quad 
\frac{1}{2}
\l1
\frac{
\sigma^3 
-
 (\sigma^3 - 3 \sigma^2 + 6 \sigma  -2) \epsilon_c
+
\frac{1}{3}
\sigma (\sigma^2 - 4 \sigma  +6) \epsilon_c^2
}
{
(\sigma - \epsilon_c)^2
}
+ 
{\cal O}
  \left(\l1^2,\,\l1_\gamma^2,\,\l1 \l1_\gamma
  \right)
,
\lb{e-approx2}
\ee
\ew
and it is implied that
$
\Theta_q \ll 1/(M n)
,
$
of course.
The latter condition obviously fails when
%$\epsilon_c - 1$ and
$\sigma$ approaches $\epsilon_c$.
%The former resonance never happens once we have assumed the Cherenkov wave's energy being less than the vacuum one. 
The  relations $\sigma = \epsilon_c$ and $\epsilon_c = 1$
are the 
horizon-type resonance conditions which can be fulfilled
only when
the particles' Compton wavelengths become
comparable with $\ell_0$,
or, equivalently, when 
the particles' momentum and energy
reach
the critical values
corresponding to the above-mentioned vacuum phase transition.
In practice, however, the approximation (\ref{e-approx2}) can
cease to be valid before $\sigma$ approaches the resonance
-
due to the violation of the trivial
requirement $\cos^2\theta \leqslant 1$.
However, this can happen only when the $\sigma$-dependent factors in the 
formula (\ref{e-approx2}) overrun $\eta$'s which are assumed
to be small by construction.

%So far this formula is approximate in velocities and exact in $ \epsilon_c$ and $\sigma$, and
%using the fact that the latter two are both proportional to $\hbar$, we can simplify the correction further more
One can see also that this correction
consists of two contributions reflecting the
fact that both the Cherenkov wave and 
the emitting particle experience
the vacuum.
For practical purposes the $\Theta_q$ corrections are small 
and can be neglected in the leading-order
approximation.
The classical limit can be attained here by setting all $\eta$'s
to zero identically since, according to the formulae above,
they vanish when the corresponding 
constants of refraction of the physical vacuum do, see also 
%the paragraph after Eq. (\ref{e-vnpt}).
the Footnote 2.

\scn{Flash duration}{sec-fla}

In the non-dispersive medium the wavefront of the Cherenkov shock
is infinitely thin, therefore, the light pulse an observer sees
when the wave hits a detector has an infinitely short duration.
However, as long as our vacuum is a dispersive
medium, the cone angle is different 
for different wavelengths.
Therefore, an observer
tuned to the frequency band 
$[\omega_1, \, \omega_2]$
will see the light flash
with a finite 
duration
$
\Delta t
=
\frac{\rho}{M c}
\left(
\tan{\theta (\omega_2)}
-
\tan{\theta (\omega_1)}
\right)
,
$
where $\rho$ is a distance from the axis of particle's trajectory.
Using the expression for the cone angle derived above,
we obtain
\be
\Delta t
\approx
\frac{\rho \Gamma \sqrt{{{\omega }_0}}}{c}
  \frac{ {\sqrt{2}}
      }
     { 1 - \sigma /\epsilon_c  }
\left( 
\frac{1}{{\sqrt{{{\omega }_2}}}}
- \frac{1}{{\sqrt{{{\omega }_1}}}}   
\right)  
\left(
1 - 
\frac{{{\mu }_{\gamma }}}{4}
\right)   
,     
\ee
where we neglected terms of 
the order
${\cal O}
\left(
\mu_\gamma^{3/2}, 
\l1^{3/2}, \l1_\gamma^{3/2},
\sqrt{\omega_1/\omega_0},\sqrt{\omega_2/\omega_0}
\right)$.
The last term again
consists of two contributions - due to 
not only the Cherenkov wave but also 
the emitting particle are affected by
the vacuum.

\scn{Energy and spectral distribution}{sec-esr}

As long as the energy of the Cherenkov photon  is 
small compared to the natural vacuum energy scale 
one can treat the problem in a linearized way where
the vacuum effects are taken into account
via the nontrivial refraction index.
By doing that we are neglecting the microscopical structure
of the vacuum which makes sense as long as the frequency
of the electromagnetic wave is  smaller than
the frequency of the vacuum oscillations $\omega_0$.
In other words, as a leading approximation
we consider the  Frank-Tamm approach with
$n$ given by Eq. (\ref{e-refi}).
Following the method, we assume 
that the $\omega$-Fourier images of the vector and scalar 
potential of the electromagnetic wave emitted by a charge
$Q$ moving at the speed $v = \text{const}$ along $z$-axis 
%in the medium with the refractive index $n$
are obeying
the macroscopic Maxwell equations in the medium
\ba
&&
\left(
\nabla^2 
+
\frac{\omega^2 n^2}{c^2}
\right)
\textbf{A}_\omega
=
- \frac{2 Q}{c} 
%\textbf{j}_\omega
\text{e}^{-i \omega z/v}
\delta (x)
\delta (y)
\textbf{e}_z
,
\\&&
\left(
\nabla^2 
+
\frac{\omega^2 n^2}{c^2}
\right)
\phi_\omega
=
- \frac{4 \pi}{n^2} 
\varrho
,
\ea
where
$\omega$ is a wave frequency, 
$\varrho$ is a charge density.
The 
%potentials are related by the equation
%$\text{div} \textbf{A}_\omega=\frac{i \omega n^2}{c}\phi_\omega,$ whereas 
Fourier images of field strengths 
are given by
$\textbf{H}_\omega = \nabla \times \textbf{A}_\omega$
and
$\textbf{E}_\omega = - (1/c) \partial_t \textbf{A}_\omega - \nabla \phi_\omega$.
Introducing the cylindrical coordinates
$\rho$, $\varphi$ and $z$,
we assume the (Fourier image of) vector potential in the form
$A_\rho = A_\varphi =0$
and
$A_z = u (\rho) \text{e}^{-i \omega z/v}$
where $u (\rho)$ obeys the differential equation
$
u''(\rho)
+
\frac{1}{\rho} 
u'(\rho)
-
\kappa
u(\rho)
=
\frac{Q}{\pi c \rho}
\delta (\rho)
,
$
where
$\kappa = 
(\omega/v)^2
\left( 1 - M^2 n^2 \right)$.
This equation can be replaced by the homogeneous 
one
if we impose the singular boundary condition in the origin:
$\lim\limits_{\rho\to 0} \rho u'(\rho) = - Q/(\pi c)$.

Now, if a charge moves slower than light then $\kappa$
is positive and the solution exponentially 
decreases with $\rho$.
Otherwise the solution has an oscillating behavior
at large $\rho$,
$
A_z 
%(\omega)
\propto
-
\frac{Q}{c \sqrt{-2 \pi \kappa \rho}}
\exp{
\left[
i \omega (t - z/v)
+
i
(\kappa + 3 \pi/4)
\right]
}
,
$
which indicates the Cherenkov wave's existence.
The total energy radiated by the charge
through the cylindrical surface of length $l$
whose axis coincides with the charge's trajectory
is given by
$
W =
\frac{1}{2} 
%\pi 
c \rho l
\int\limits_{-\infty}^{\infty}
d t
\int\limits_{M n \geq 1}
d \omega
d \omega'
\text{e}^{(\omega+\omega')t}
\left|
\textbf{E}_\omega 
\times
\textbf{H}_{\omega'} 
\right|
,
$
where the integration over the frequencies
must be performed only for those values at which the
charge's velocity is larger than $c_n$ but smaller
than $c$.
According to dispersion relations,
the latter bound imposes the natural ultraviolet cut-off 
$E_0/\hbar = - \xi \omega_0 $
such that one does not need to postulate it separately,
in contrast to the conventional
materials.
By
introducing the variable $x = - \xi \omega/ \omega_0$
we can do both cases $\xi = \pm 1$ in a uniform way.
The radiation energy per unit path
is given by the 
Frank-Tamm  
spectral distribution
\be
\frac{d W}{d l}
=
\frac{\mu_\gamma Q^2 \omega_0^2}{c^2}
\int\limits_{0}^{1}
d x
\frac{
x (x-1)^2
}
{
\mu_\gamma (x-1)^2
+1
}
+
{\cal O} (\l1, \hbar)
,
\ee
with the integrand having a local maximum
at $x_\text{peak} = 1/3 + {\cal O} (\mu_\gamma)$.
Thus,
% that we obtain that
the radiation yield produced by the moving charge 
$Q$ 
%moving with the constant velocity $v$, $c_n \leqslant v \leqslant c$, 
amounts to
\[
\frac{d W}{d l}
=
\frac{\mu_\gamma}{3}
\left(\frac{e N E_0}{2 c \hbar}\right)^2
%(9 + 8\xi)
+
{\cal O} (\l1, \hbar, \mu_\gamma^2)
,
\]
where $N = Q/e $, $e$ being the elementary charge.
Therefore, 
%for, e.g., $\xi = -1$ 
in the leading-order approximation 
we obtain
\be
\frac{d W}{d l}
=
3 \times 10^{10} \mu_\gamma N^2 E_0^2 \ \text{GeV}^{-1} \text{cm}^{-1}
.
\ee
Looking at these equations we can immediately notice that 
the value of the vacuum energy enters the picture
in a crucial way.
As a matter of fact, the 
%appearance of 
main contributing prefactor, 
%$Q^2 \omega_0^2 \sim \frac{N^2 E_0^2}{\hbar c}$,
$\omega_0^2 \sim E_0^2$,
is inherent in the theory of the Cherenkov radiation in 
effectively continuous media.
Moreover,
its appearance
weakly depends on a specific form of the refractive index - as long
as the latter contains the ultraviolet cut-off frequency.
Therefore, this factor should
necessarily appear in a very large class of theories 
with the  
ultraviolet cutoff being
determined by the energy of the vacuum.
The non-local nature of the superfluid vacuum
as an extended object
(i.e., non-point-like and possessing internal structure)
makes the quantitative properties of the Cherenkov effect in theories with the
BEC-type vacuum being different from those in some
other LIV theories.
Our results are
more close to the predictions based on the general arguments  
about the existence of a preferred frame of reference \cite{Coleman:1997xq} -  
which is not surprising though.
%The relativistic superluminal particles (tachyons) can not emit Cherenkov radiation
%due to the trivial structure of vacuum \cite{Mignani:1973pq}.

The radiation yield of the Cherenkov effect in the
usual materials is observed to be relatively small,
few keV per cm,
but only
because the typical ultraviolet cutoff
frequency there is tiny small - about 
an electronvolt per Planck.
However, on a scale of the cutoff frequency the energy output is 
not small at all -
it is 
%experimentally observed to be 
at least three orders of magnitude
larger than the cutoff energy:
$(d W/ d l)_{N=1}  \sim E_0 \times 10^3 \, \text{cm}^{-1}$.
%per particle) 
In vacuum 
%Cherenkov effect 
the cutoff energy
is higher by many orders of magnitude
and also
may depend on a physical setup
because the background condensate gets affected by 
geometry of the problem and external fields acting upon
which leads to the value of $E_0$ can
differ for different physical situations
(same goes about other parameters such as $\mu$'s).
Therefore,
in absence of the proper microscopical theory 
of the physical vacuum, the value 
$E_0$ is difficult to compute theoretically,
yet the boundaries can be established already at this stage.
%in a heuristic way.
The upper boundary for $|E_0|$ is, of course,
the Planck energy,
$10^{19}\ \text{GeV}$,
the largest energy pertinent to the microworld
(debates, however, continue \cite{Gagnon:2004xh}). 
The lower boundary, $10^4 \ \text{GeV}$, comes 
from current non-observability data,
and thus can be significantly lifted \cite{Klinkhamer:2005cp}.
Using these  conservative values,
%for the vacuum energy, 
$10^4  \lesssim |E_0| \lesssim 10^{19}\ \text{GeV} $,
we give the following estimate 
%the bounds 
%for the radiation yield the following bounds:
\be
10^{15} 
\lesssim
\frac{1}{\mu_\gamma N^2} 
\frac{d W}{d l}
\lesssim
10^{45} \ \text{TeV/cm}
.
\ee
%Of course, these numbers will be significantly dumped by 
Despite the constant of refraction of the vacuum is obviously extremely small,
the resulting numbers can  be quite substantial
-
especially considering that $N$ can be large.
In that case
the vacuum Cherenkov shocks turn out to be a
very efficient, fast and powerful
way of 
draining and releasing energy.
This poses the question whether such processes 
can happen in the astrophysical objects such as
super- and hypernovas \cite{Paragi:2010tu,Soderberg:2009ps},
active galactic nuclei, gamma-ray bursts
and ROCOSs (radio objects with the continuous optical spectra often
having an abnormally strong ultraviolet part  \cite{rocos}).

\scn{Beyond Frank-Tamm: ``Boom shock''}{sec-bft}

In a conventional theory 
of the Cherenkov effect the Frank-Tamm formula
was derived assuming that the particle's velocity
is constant and any changes of it happen 
instantaneously.
This approximation has been proven to be very robust,
yet in reality the speed-of-light barrier
is crossed by the particle which is either
accelerating or decelerating in a smooth way.
The analytical theory of the Cherenkov radiation
for such cases is far from being complete, even 
for the case of conventional materials. 
There exists, however, a number of heuristic and numerical results
which point at the appearance of the separate wave
when the velocity of a moving charge exactly coincides
with the
speed of light in the medium - 
the so-called Tyapkin-Zrelov(-Afanasiev) or ``luminal boom'' 
wave \cite{tza,tza2}.
In the case of an accelerating charge such wave
is indistinguishable from the Cherenkov one as
it just closes the cone but
for the decelerating motion this wave decouples 
from the charge when the latter crosses the $c_n$-barrier
while slowing down.
Then this wave
continues propagating independently with the velocity $c_n$.

\scn{Conclusion}{sec-cnc}

%To conclude,
In this paper we  theoretically outlined basic properties
of the 
%radiation phenomena in vacuum, including the 
Cherenkov-type radiation phenomena in vacuum.
%and Hawking ones. 
It is shown that
the macroscopical description
of the Cherenkov radiation is based on two parameters, 
the constant of refraction and the cut-off frequency.
% which is defined by the vacuum energy.
From the phenomenological point of view, even 
in the conventional materials such parameters
are quite difficult to determine
theoretically but 
the experimental findings are greatly facilitated
by 
the universal features 
%and material-independence
of the Cherenkov radiation.
The same is true for the vacuum case,
moreover, 
when
compared to other possible dissipative processes
then in terms of released energy  
the Cherenkov effect in vacuum
should play more dominant role 
than its 
%material 
condensed-matter
counterpart (the latter usually accounts for less than one
per cent of the energy loss by ionization).
This becomes possible 
because the vacuum itself can be viewed as the (super)fluid
with minimum dissipation.

While our study was mainly based on the theory described
by the logarithmic nonlinear Schr\"odinger equation,
some of the obtained results 
must be valid
for any Lorentz-invariance violating theory 
describing the 
vacuum by (effectively) continuous medium in the
long-wavelength approximation.
%Besides, here it will help a lot that the studies on the Earth's accelerators are being supplemented by astrophysical observations.

\begin{acknowledgments}
I am grateful to Eugene Tkalya and Ralf Lehnert
%from Moscow State University
for bringing some references into my attention and fruitful comments.
%Useful correspondence from  is acknowledged.
This work was supported under a grant of the National Research Foundation of South Africa. 
\end{acknowledgments}

\def\AnP{Ann. Phys.}
\def\APP{Acta Phys. Polon.}
\def\CJP{Czech. J. Phys.}
\def\CMPh{Commun. Math. Phys.}
\def\CQG {Class. Quantum Grav.}
\def\EPL  {Europhys. Lett.}
\def\IJMP  {Int. J. Mod. Phys.}
\def\JMP{J. Math. Phys.}
\def\JPh{J. Phys.}
\def\FP{Fortschr. Phys.}
\def\GRG {Gen. Relativ. Gravit.}
\def\GC {Gravit. Cosmol.}
\def\LMPh {Lett. Math. Phys.}
\def\MPL  {Mod. Phys. Lett.}
\def\Nat {Nature}
\def\NCim {Nuovo Cimento}
\def\NPh  {Nucl. Phys.}
\def\PhE  {Phys.Essays}
\def\PhL  {Phys. Lett.}
\def\PhR  {Phys. Rev.}
\def\PhRL {Phys. Rev. Lett.}
\def\PhRp {Phys. Rept.}
\def\RMP  {Rev. Mod. Phys.}
\def\TMF {Teor. Mat. Fiz.}
\def\prp {report}
\def\Prp {Report}

\def\jn#1#2#3#4#5{{#1}{#2} {\bf #3}, {#4} {(#5)}} %PRD
%\def\jn#1#2#3#4#5{{#1}{#2} {#3} {(#5)} {#4}}   %PLB style
% #1 tittle  #2 ser  #3 vol  #4 page  #5 year

\def\boo#1#2#3#4#5{{\it #1} ({#2}, {#3}, {#4}){#5}}
%\def\boo#1#2#3#4#5{ #1 ({#2}, {#3}, {#4}){#5}}  %PLB style
% #1 tittle  #2 publisher  #3 place  #4 year  #5 page/, p.789/

%\def\jn#1#2#3#4#5{{#1}{#2} {\bf #3}, {#4} {(#5)}}
% #1 tittle  #2 ser  #3 vol  #4 page  #5 year
%\def\boo#1#2#3#4#5{{\it #1} ({#2}, {#3}, {#4}){#5}}
% #1 tittle  #2 publisher  #3 place  #4 year  #5 page/, p.789/

%\newpage

\end{document}